\documentclass[%
 reprint,
 amsmath,amssymb,
 aps,
]{revtex4-1}

\usepackage{graphicx}
\usepackage{dcolumn}
\usepackage{bm}

\begin{document}

\preprint{APS/123-QED}

\title{Invariance of  Bipartite Separability and PPT-Probabilities over Casimir Invariants of Reduced States}

\author{Paul B. Slater}
\affiliation{Kavli Institute for Theoretical Physics, University of California, Santa Barbara.}
\email{slater@kitp.ucsb.edu}

\date{\today}

\begin{abstract}
Milz and Strunz ({\it J. Phys. A}: {\bf{48}}  [2015] 035306) recently studied the probabilities that two-qubit and qubit-qutrit states, randomly generated with respect to Hilbert-Schmidt (Euclidean/flat) measure, are separable. They concluded that in both  cases, the separability probabilities 
(apparently exactly $\frac{8}{33}$  in the two-qubit scenario) hold {\it constant} over the Bloch radii
($r$) of the single-qubit subsystems, jumping to 1 at the pure state boundaries ($r=1$). 
Here, firstly, we present evidence that in the qubit-qutrit case, the separability probability is uniformly distributed, as well, over the {\it generalized} Bloch radius ($R$) of the qutrit subsystem. While the qubit (standard) Bloch vector is positioned in three-dimensional space, the qutrit generalized Bloch vector lives in eight-dimensional space. The radii variables $r$ and $R$ themselves are the lengths/norms (being square roots of {\it quadratic} Casimir invariants) of these  (``coherence'') vectors.  Additionally, we find that not only are the qubit-qutrit separability probabilities invariant over the quadratic Casimir invariant of the qutrit subsystem, but apparently also over the {\it cubic} one--and similarly the case, more generally,  with the use of random induced measure.  We also investigate two-qutrit ($3 \times 3$) and qubit-{\it qudit} ($2 \times 4$) systems--with seemingly analogous {\it positive-partial-transpose}-probability invariances holding over what have been termed by Altafini, the {\it partial} Casimir invariants  of these systems.
\end{abstract}

\maketitle

\tableofcontents
\section{Introduction}
The separable quantum states are embedded in the set of all (separable and entangled) states \cite{Gamel}. The nature of this embedding and its magnitude, in terms of various measures of quantum-theoretic interest \cite{ingemarkarol,petzsudar}, are fundamental (``philosophical,\ldots,practical,\ldots,physical" \cite{ZHSL})  issues \cite{ingemarkarol,sbz}. We present below apparent relations between these issues and Casimir invariants \cite{Gerdt2011}---distinguished elements of the center of the universal enveloping algebra of  a Lie algebra. In the simplest, lowest-dimensional case exhibiting entanglement, that of two qubits, this invariant is the square of the  familiar Bloch radius of either qubit subsystem. For a system with a three-state (qutrit) subsystem, one of the two Casimir 
invariants is the square of the corresponding ``generalized Bloch radius" \cite{Goyal,Kimura,scutaru2}, and similarly for higher-dimensional (qudit) subsystems.

In such regards, a diverse body of evidence--though yet no formal proof--has been developed, strongly indicating that the probability that a two-qubit state is separable/disentangled/classically correlated, 
that is, expressible as the convex sum of products of qubit states \cite{ClassicallyCorrelated}, is 
$\frac{8}{33} \approx 0.242424$ \cite{slater833,MomentBased,slaterJModPhys,FeiJoynt}. 
The measure employed in the underlying computations was the familiar Hilbert-Schmidt (Euclidean/flat) one \cite{szHS,ingemarkarol}, while the integration of this measure was conducted over the standard 15-dimensional convex set of $4 \times 4$ (Hermitian) density matrices. (The separability probability is computed as the ratio of the Hilbert-Schmidt volume of separable states to the volume of all states \cite{ZHSL}.)

Let us also note--though they will not be further discussed here--that still other simple exact rational-valued separability probabilities appear to hold in related scenarios, with the use of Hilbert-Schmidt measure--as well as its generalization to {\it random-induced} measures \cite{Induced}. 
Notable examples are
the 9-dimensional two-re[al]bit and 27-dimensional two-quater[nionic]bit density matrices.

Relatedly, a  ``concise'' infinite summation formula 
\begin{equation} \label{Hou1}
P(\alpha) =\Sigma_{i=0}^\infty f(\alpha+i),
\end{equation}
where
\begin{equation} \label{Hou2}
f(\alpha) = P(\alpha)-P(\alpha +1) = 
\end{equation}
\begin{displaymath}
\frac{ q(\alpha) 2^{-4 \alpha -6} \Gamma{(3 \alpha +\frac{5}{2})} \Gamma{(5 \alpha +2})}{3 \Gamma{(\alpha +1)} \Gamma{(2 \alpha +3)} 
\Gamma{(5 \alpha +\frac{13}{2})}},
\end{displaymath}
and
\begin{equation} \label{Hou3}
q(\alpha) = 185000 \alpha ^5+779750 \alpha ^4+1289125 \alpha ^3
\end{equation}
\begin{displaymath}
+1042015 \alpha ^2+410694 \alpha +63000,
\end{displaymath}
appears to apply in the Hilbert-Schmidt instances \cite{slaterJModPhys}.  (The formula was constructed through an application by Qing-Hu Hou of  the famous procedure of ``creative telescoping'' of Doron Zeilberger \cite{doron} to a lengthy hypergeometric-based expression.) Here, 
$\alpha$ functions as a Dyson-index-like parameter of random matrix theory (cf. \cite{MatrixModels}). The formula yields
$P(\frac{1}{2}) =\frac{29}{64}$ in the two-rebit case, $P(2)= \frac{26}{323}$ in the 
two-quaterbit scenario, as well as (apparently even more simply) the mentioned $P(1)= \frac{8}{33}$ in the (standard) two-qubit case \cite{FeiJoynt}.
\subsection{Contribution of Milz and Strunz}
A further interesting contribution to this general area of separability-probability research (originating in the seminal paper of {\.Z}yczkowski, Horodecki, Sanpera and Lewenstein \cite{ZHSL}) was recently made by Milz and Strunz \cite{milzstrunz}. They studied cases of random (with respect to Hilbert-Schmidt measure) $2 \times n$ ($n=2, 3, 4$) Hermitian density matrices. 
They found evidence that the putative (overall) separability probability of 
$\frac{8}{33}$ appeared 
remarkably to hold {\it constant} along the Bloch radii ($r$) of the qubit subsystems in the $n=2$ case \cite[eq. (31)]{milzstrunz}, and also constant
(but with smaller probabilities--cf. \cite[eqs. (3)-(5)]{LatestCollaboration}--in the $n = 3, 4$ cases). In the $n=4$ qubit-{\it qudit} setting, the probability  employed was that of having a positive partial transpose (PPT). (These uniformities do appear to hold in the half-open interval $r \in [0, 1)$ , jumping to 1 at the pure state boundary, that is, $r=1$.) 

``The Bloch sphere provides a simple representation for the state space of the most primitive 
quantum unit--the qubit--resulting in geometric intuitions that are invaluable in countless fundamental information-processing scenarios'' \cite{Jevtic}.
\subsection{Repulsion phenomenon in joint two-qubit separability probabilities}
Motivated by this recent work of Milz and Strunz, we were led to examine \cite{Repulsion} in the specific $n=2$ 
two-qubit case the nature of the 
{\it bivariate} (joint) separability probability over the pair of Bloch radii ($r_A, r_B$)--that is, the norms/lengths of the 
Bloch/coherence \cite{PhysRevA.68.062322} vectors 
of the induced 
single-qubit subsystems ($A,B$). A certain {\it repulsion} phenomenon was uncovered.

That is, separability probabilities tended to be smaller, the closer in length that their two Bloch radii were to each other. 
(It appears to be an interesting research question of in what manner such observations are related to findings, pertaining to the use of the Ky Fan norm, in \cite{DeVicente}. There, de Vicente asserts that ``Theorem 1 has a clear physical meaning: there is an {\it upper bound} (emphasis added) to the correlations contained in a separable state''.)  The exact nature of the (now, clearly nonuniform) bivariate 
distribution over the pair of Bloch radii \cite[Fig. 5]{Repulsion}, however, remains 
to be determined in this two-qubit and related (real, quaternionic, induced measure,\ldots) cases.
\section{Qubit-Qutrit analysis} \label{2x3A}
We begin our series of analyses here by further examining the qubit-qutrit case, that is the $2 \times n$ scenario with $n=3$. Presumably, by the analyses of Milz and Strunz \cite[Fig. 4]{milzstrunz}, the Hilbert-Schmidt separability probability holds constant over the Bloch radius ($r$) 
of the  single-qubit subsystem. 

Now, we investigate, additionally, the variation of the separability probability over the generalized Bloch radius  $R$ 
of the induced single-{\it qutrit} subsystem 
\cite[eqs. (7), (15)]{Goyal} \cite{Kimura,scutaru2} (the Bloch vector now being situated in 8-dimensional space) .
(``In place of the three Pauli matrices we now need the \ldots eight Gell-Mann $\lambda$-matrices to describe a generalization of the Bloch ball representation of qubit to the case of three-level system or qutrit \ldots
These matrices are familiar as generators of the unimodular unitary group SU(3) in its defining representation. Just like the Pauli matrices these form a complete set of hermitian, traceless, trace-orthogonal matrices'' \cite{Goyal}.)

Accordingly, we generated $N_{tot} =10^8$  (one hundred million) qubit-qutrit density matrices, randomly with respect to Hilbert-Schmidt measure, employing the Ginibre ensemble methodology 
\cite[eq. (15)]{generating}. Each such state was tested for its separability--that is, whether or not the six eigenvalues of the  
{\it partial transpose} (PT) of the density matrix were all nonnegative \cite{asher,michal}. The Bloch radii 
($r$ and $R$) were found for its reduced single-spin qubit and qutrit subsystems.

The number of separable density matrices found was 
$N_{sep}=2,699,590$, with
the  qubit-qutrit separability probability estimate accordingly being 0.0269959. We note that the associated $99.9\%$ confidence interval of 
$\{0.0269426,0.0270492 \}$ does {\it not} include a previously conjectured value of
$\frac{32}{1199} \approx 0.026688 $ \cite[sec. 10.2]{slater833}. (This conjecture had been arrived at in 2007 with the use of quasi-Monte Carlo sampling methods, rather than the now preferable Ginibre ensemble approach \cite[eq. (15)]{generating}, employed in this current study.) Milz and Strunz themselves did report an estimate of $0.02700 \pm 0.00016$ \cite[eq. (33)]{milzstrunz}. (We report a second, independent estimate--which we pool with this one--in sec.~\ref{2x3B}.)

The values recorded of $r$ and $R$, ranging from zero to one (having been  appropriately scaled in the qutrit case \cite[eq. 6]{PhysRevA.68.062322} 
\cite[eq. (12)]{Goyal}), were discretized
into intervals of length $\frac{1}{100}$. Thus, we generated two
data matrices of dimensions $100 \times 100$, one corresponding to the one hundred million random $6 \times 6$ density matrices generated, and one for the subset of separable density matrices.

\begin{figure}
\includegraphics[scale=0.65]{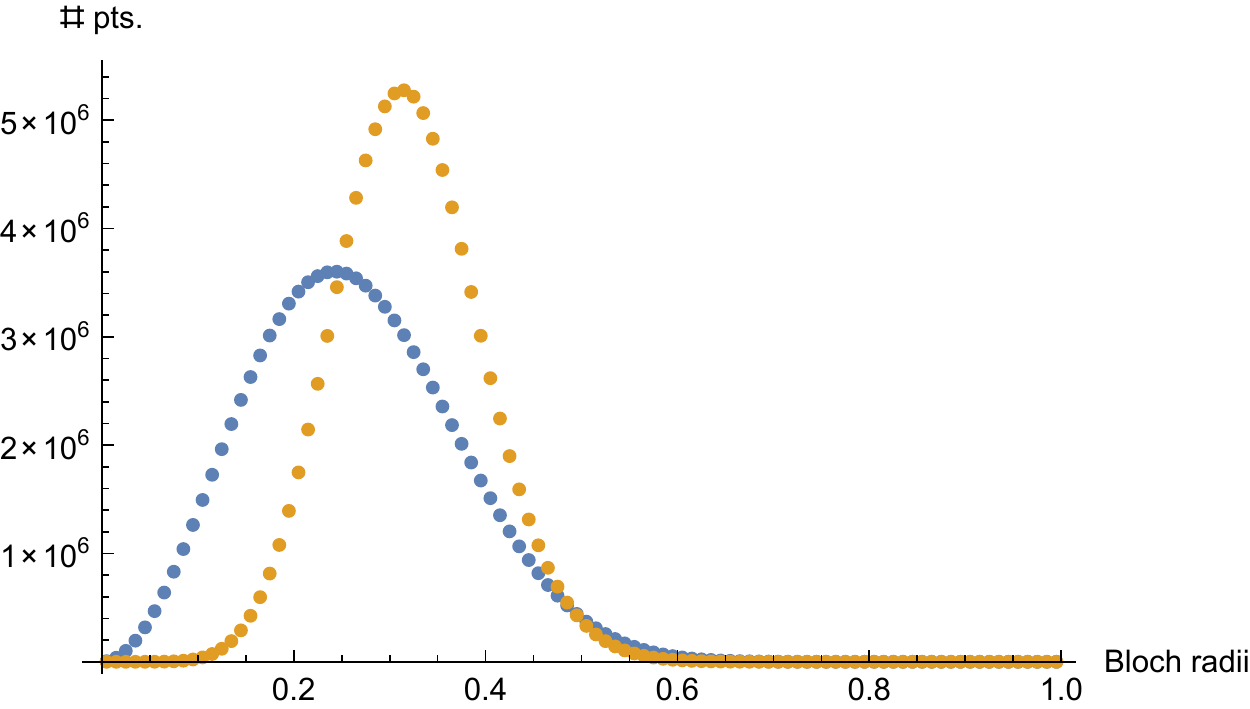}
\caption{\label{fig:Allstates}Distributions (histograms) of sampled qubit-qutrit states over the Bloch radii, with the more sharply-peaked distribution corresponding to the qutrit generalized Bloch radius $R$.}
\end{figure}

\begin{figure}
\includegraphics[scale=0.65]{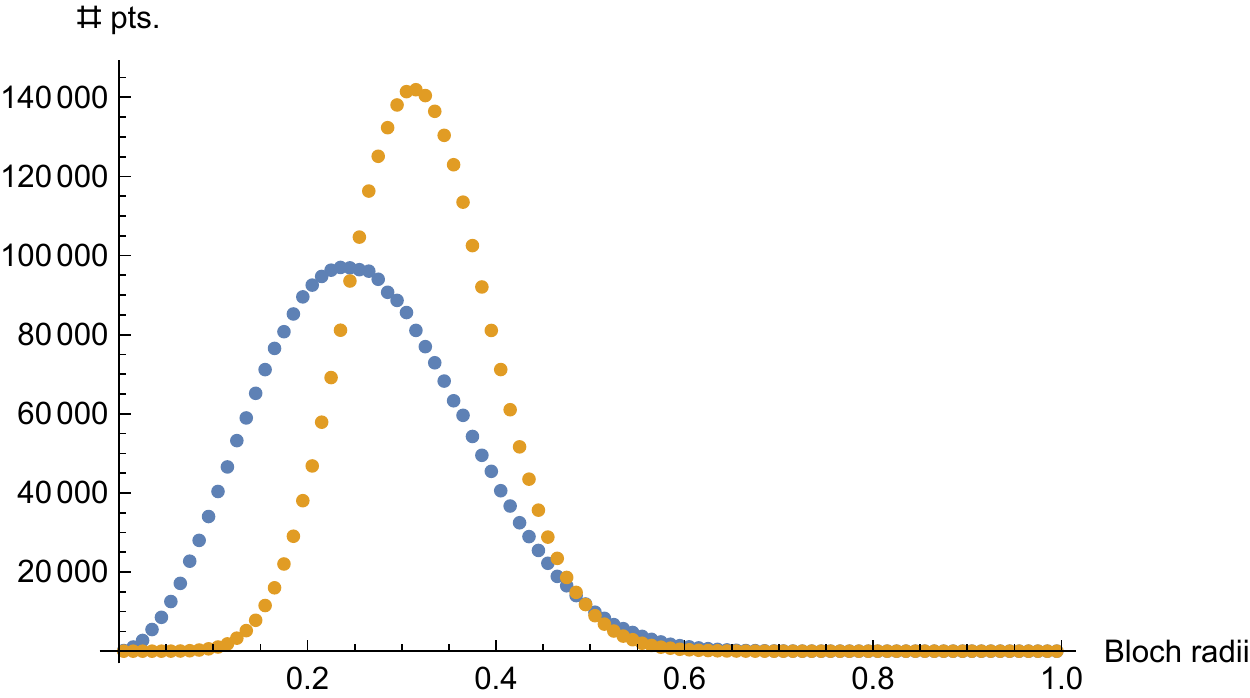}
\caption{\label{fig:Separablestates}Distributions of sampled 
{\it separable} qubit-qutrit states over the Bloch radii, with the more 
sharply-peaked distribution corresponding to the qutrit generalized Bloch radius $R$.}
\end{figure}

In Fig.~\ref{fig:Allstates} 
we show the distributions (that is, the numbers recorded) of the $N_{tot}$ sampled states with respect to each of the 
Bloch radii, and in 
Fig.~\ref{fig:Separablestates}, 
similarly only the $N_{sep}$ separable states. The distribution over the qutrit radial variable $R$ is more sharply peaked in each instance--and, of course and highly importantly, we note the 
very strong similarity in distributional shapes between these first two (total and separable) figures. 
\subsection{Modeling of the qutrit $R$-curves}
Milz and Strunz 
\cite[Fig. 3, eq. (27)]{milzstrunz} conjectured that the (qubit) $r$-curve in 
Fig.~\ref{fig:Allstates} would be proportional to 
$r^2 (1-r^2)^{16}$ 
(as well as $r^2 (1-r^2)^{2 (m^2-1)}$, more generally for 
$2 \times m$ systems)--and their proposal was very well supported by our corresponding plot. 

\begin{figure}
\includegraphics[scale=0.65]{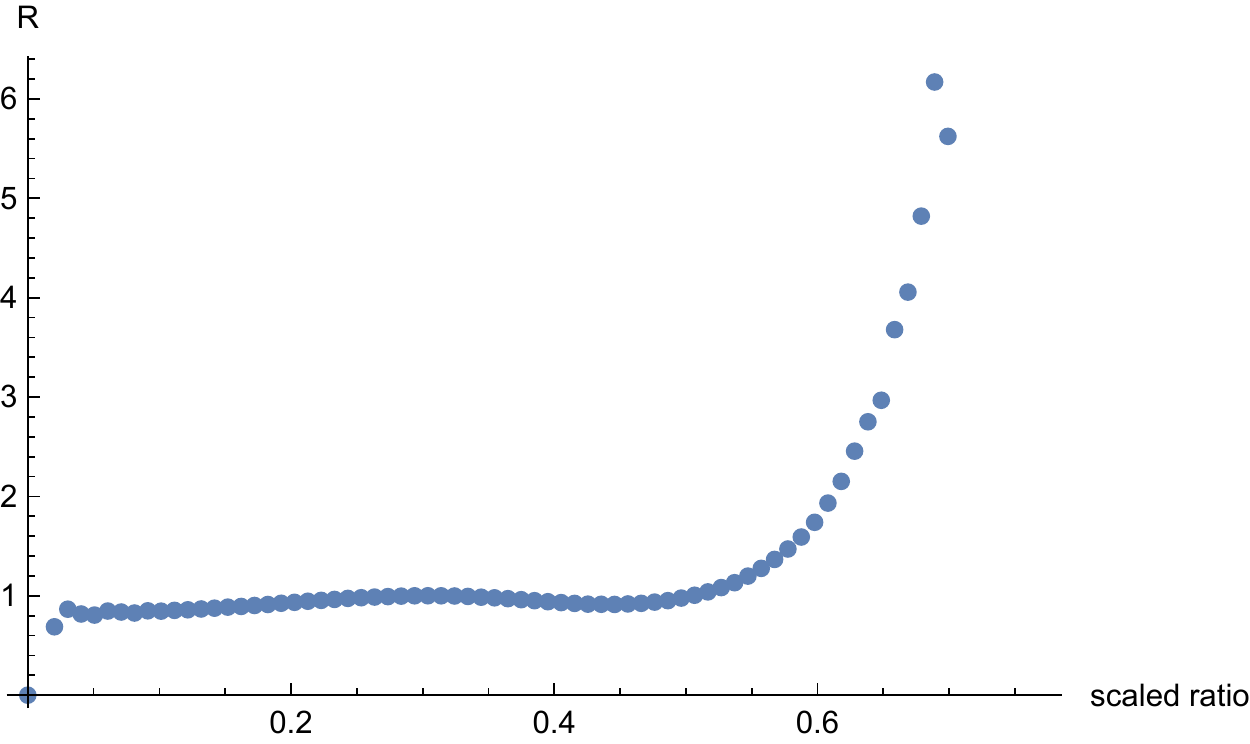}
\caption{\label{fig:ScaledRatio}Scaled ratio of the $R$-curve (for the qutrit subsystem) in Fig.~\ref{fig:Allstates} to  $R^7 (1-R^2)^{32}$.}
\end{figure}

We, now, attempted a comparable fit to the  
$R$-curve in Fig.~\ref{fig:Allstates} and found that a scaled version of 
$R^7 (1-R^2)^{32}$ succeeded fairly well over the {\it half} interval 
$R \in [0,\frac{1}{2}]$ (Fig.~\ref{fig:ScaledRatio}). 
(``Thus the boundary [of the spin-1 states] can never stray into the interior of the eight-dimensional solid sphere of radius 1/2 contained in [the spin-1 states]'' \cite[p. 4]{Goyal}.)
\subsection{$r-$ {\it and} $R$-invariances of separability probabilities}
\begin{figure}
\includegraphics[scale=0.65]{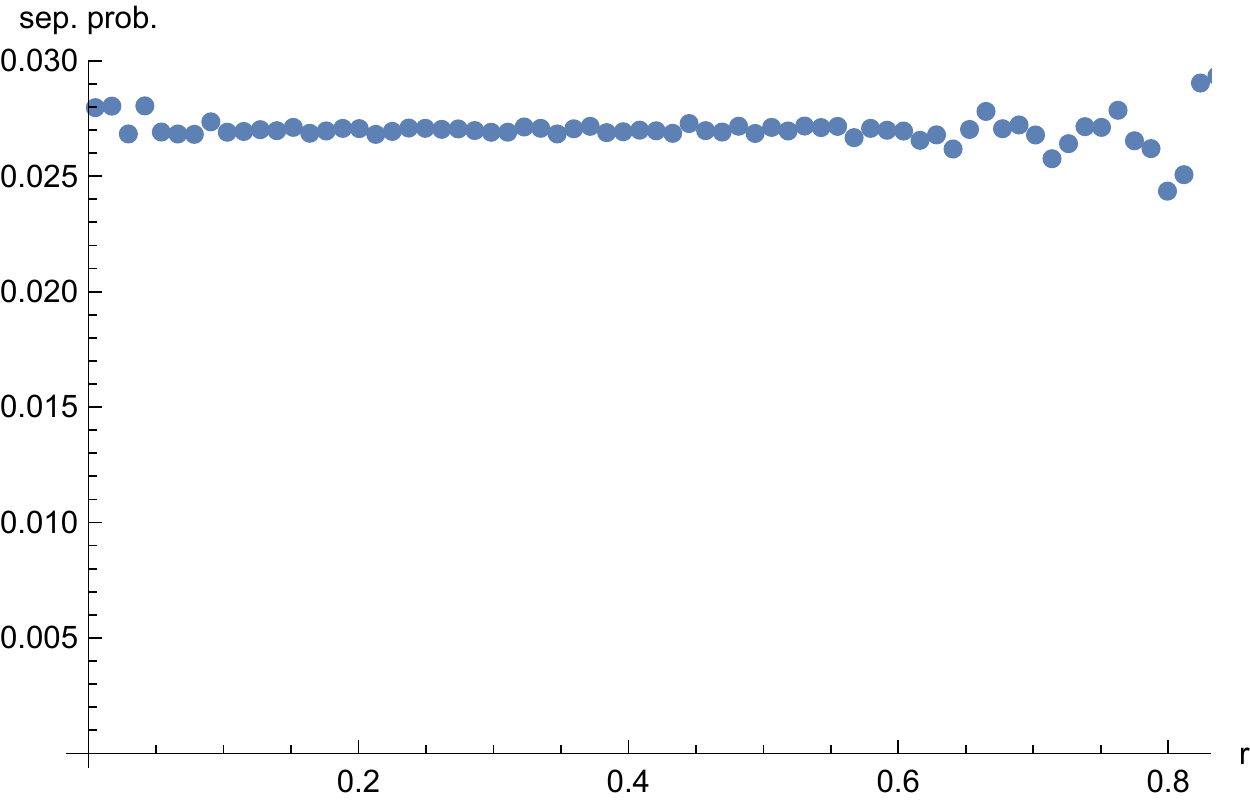}
\caption{\label{fig:Qubitprobs}Separability probability estimates given by the ratio of the (qubit) $r$-curve in 
Fig.~\ref{fig:Separablestates}  to the (qubit) $r$-curve in  Fig.~\ref{fig:Allstates}. The flatness accords
with the findings of Milz and Strunz \cite[Fig. 4]{milzstrunz} (also \cite[Fig. 10]{Repulsion}).}
\end{figure}

We take the ratios of the number of sampled separable states $N_{sep}$ to the number of  all sampled states $N_{tot}$ in both ($r, R$) cases for each subinterval of length $\frac{1}{100}$, giving us the desired {\it univariate} separability probability estimates over the pair of [0,1]  intervals.
In Fig.~\ref{fig:Qubitprobs} 
we show the counterpart to Fig. 5 in \cite{milzstrunz}, manifesting the same constancy/invariance over $r$ as observed by Milz and Strunz (which served as the initial motivation for our further study here and in \cite{Repulsion}).

\begin{figure}
\includegraphics[scale=0.65]{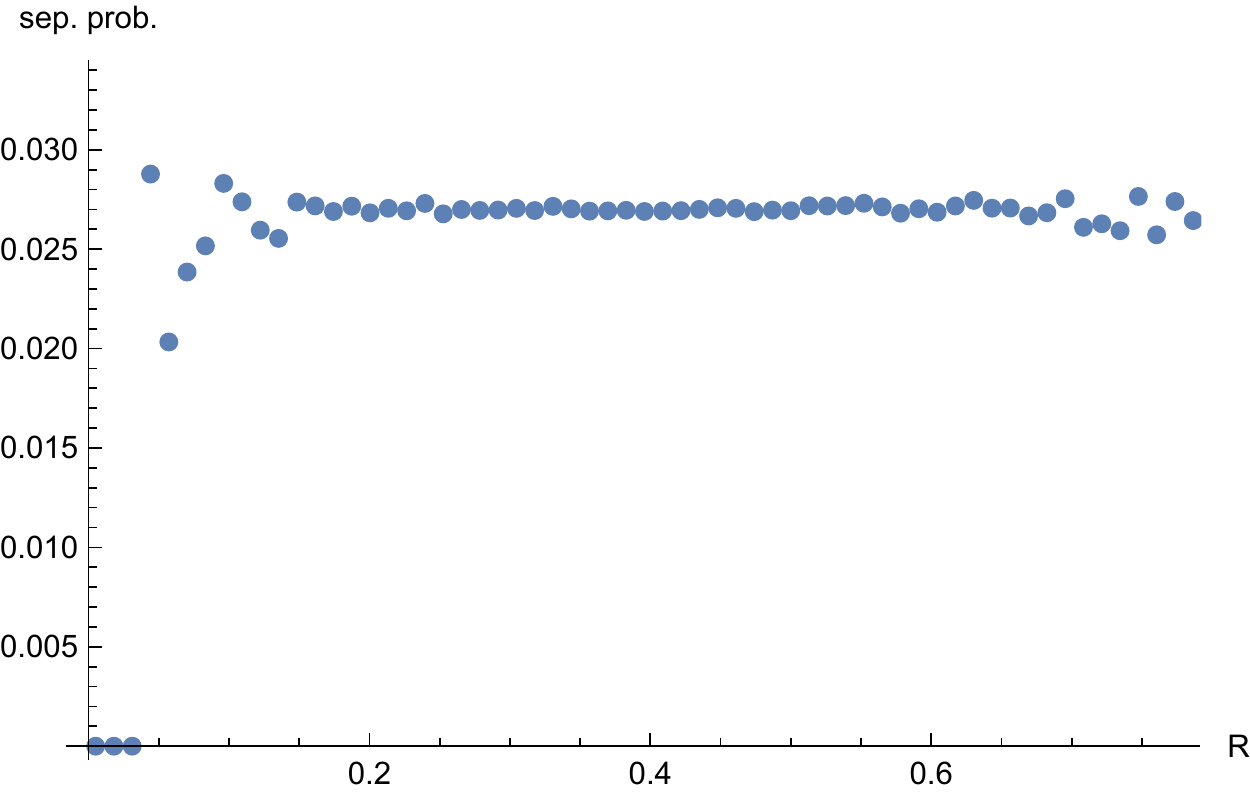}
\caption{\label{fig:Qutritprobs}Separability probability estimates given by the ratio of the (qutrit) $R$-curve in 
 Fig.~\ref{fig:Separablestates}  to the 
(qutrit) $R$-curve in  Fig.~\ref{fig:Allstates}.}
\end{figure}

Now, we newly present the $R$-counterpart 
(Fig.~\ref{fig:Qutritprobs}) to Fig.~\ref{fig:Qubitprobs},
being essentially indistinguishable in its flat character.
(We can see from 
Figs.~\ref{fig:Allstates} and \ref{fig:Separablestates} that the number of sampled qubit-qutrit states declines in both tails of the distributions, leading 
naturally to more scatter in the tails of  the two flat-like separability probability figures. It would be of interest to incorporate confidence intervals into these and certain of the succeeding figures--as employed in \cite{Repulsion}. Let us note the availability of formal statistical tests for the 
equality of a collection of binomial proportions \cite{KP}.)

Thus, it now strongly appears that the qubit-qutrit Hilbert-Schmidt 
separability probabilities hold constant (except at the pure states), not only over the 
the qubit (standard) Bloch radius $r$, as Milz and Strunz interestingly indicated, but {\it also}  over the qutrit 
{\it generalized} Bloch radius $R$. 
These parallel results are somewhat intuitive, given our first two plots (Figs.~\ref{fig:Allstates} and \ref{fig:Separablestates}), since the {\it shapes} of the two curves in both plots appear essentially identical to one another. 
\subsection{Joint qubit-qutrit separability probability}
\begin{figure}
\includegraphics[scale=0.65]{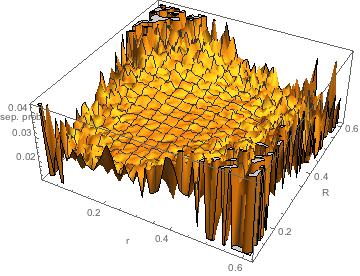}
\caption{\label{fig:JointQubitQutritPlot}Joint qubit-qutrit separability probability estimate over the qubit Bloch radius $r$ and the 
qutrit Bloch radius $R$.}
\end{figure}

In Fig.~\ref{fig:JointQubitQutritPlot}, we show our estimate of the {\it bivariate} (joint) qubit-qutrit separability probability distribution 
(cf. \cite[Fig. 5]{Repulsion} for the two-qubit counterpart).
\section{Higher-dimensional analyses}
It is of obvious interest to extend this form of analysis
to further $m \times n$ systems, where now $m n>6$, using the corresponding $(m^2-1)$ and ($n^2-1$)-dimensional forms of Bloch (coherence) vectors \cite{Kimura}. In such analyses, it would appear that the concept of positive partial transpose (PPT) probability is the appropriate one to  replace that of separability probability. Conjecturally, then, such PPT-probabilities would continue to be found to hold constant along the (generalized) Bloch radii of the induced subsystems. 
This will appear to be the case in the further analyses below.
\subsection{Two-{\it qutrit} analysis}
\begin{figure}
\includegraphics[scale=0.65]{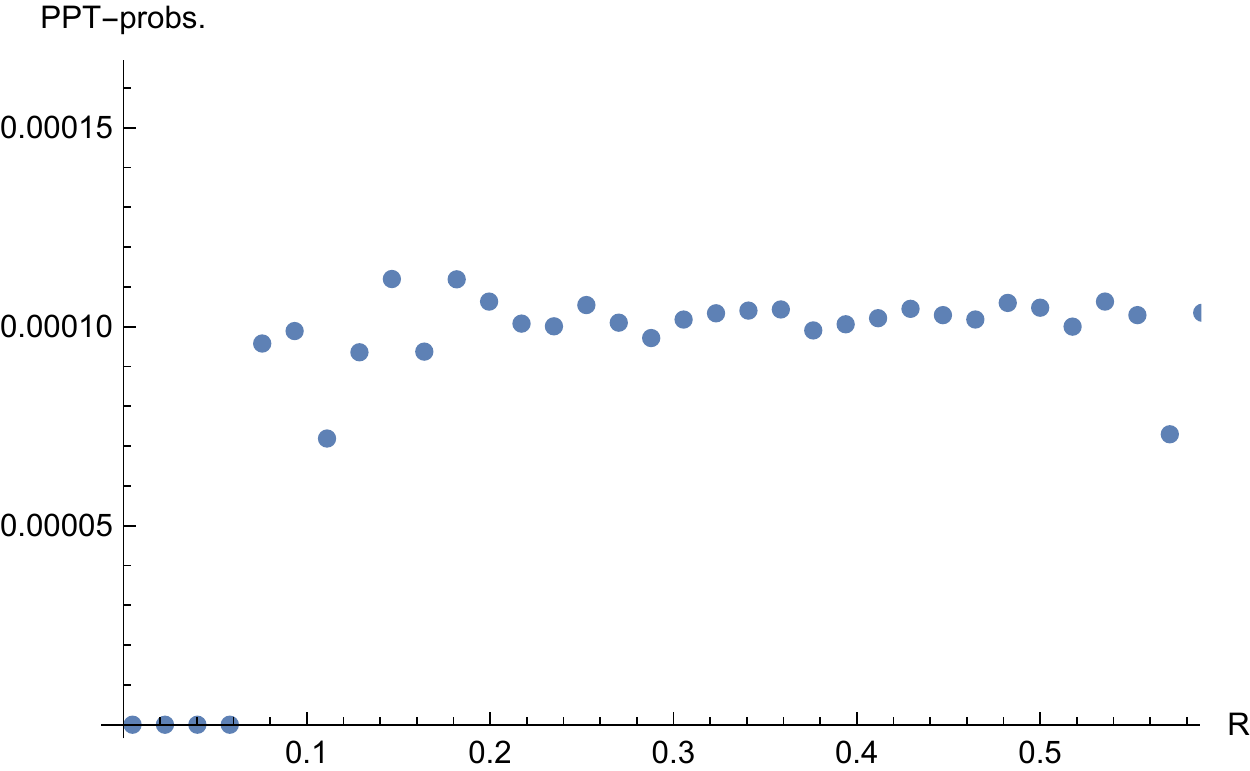}
\caption{\label{fig:TwoQutritsprob}Two-{\it qutrit} positive-partial-transpose probability estimates given by the ratio of  Fig.~\ref{fig:PPTtwoqutrits} 
to Fig.~\ref{fig:Alltwoqutrits}.}
\end{figure}

We generated 100  million $9 \times 9$ density matrices, once again of a random nature with respect to Hilbert-Schmidt measure. Regarding them as two-{\it qutrit} 
systems (cf. \cite{VertesiBrunner,Jarvis}), (only) 10,218  of them had positive partial transposes (PPT), leading to an associated PPT-probability of 
0.00010218. (Having a PPT is now a necessary, but not sufficient, condition for separability. The 95$\%$ confidence interval for the true probability was 
 $\{0.000100199,0.000104161\}$ [cf. \cite{Brown}].) We plotted these PPT-probabilities as a function of $R$ 
(Fig.~\ref{fig:TwoQutritsprob})--presumably the function is of the same nature for a choice of $R=R_A$ or $R_B$--again 
having divided the interval $R \in [0,1]$ into one hundred bins. (We had initially symmetrized the underlying $100 \times 100$ data matrix for added stability.) 

\begin{figure}
\includegraphics[scale=0.65]{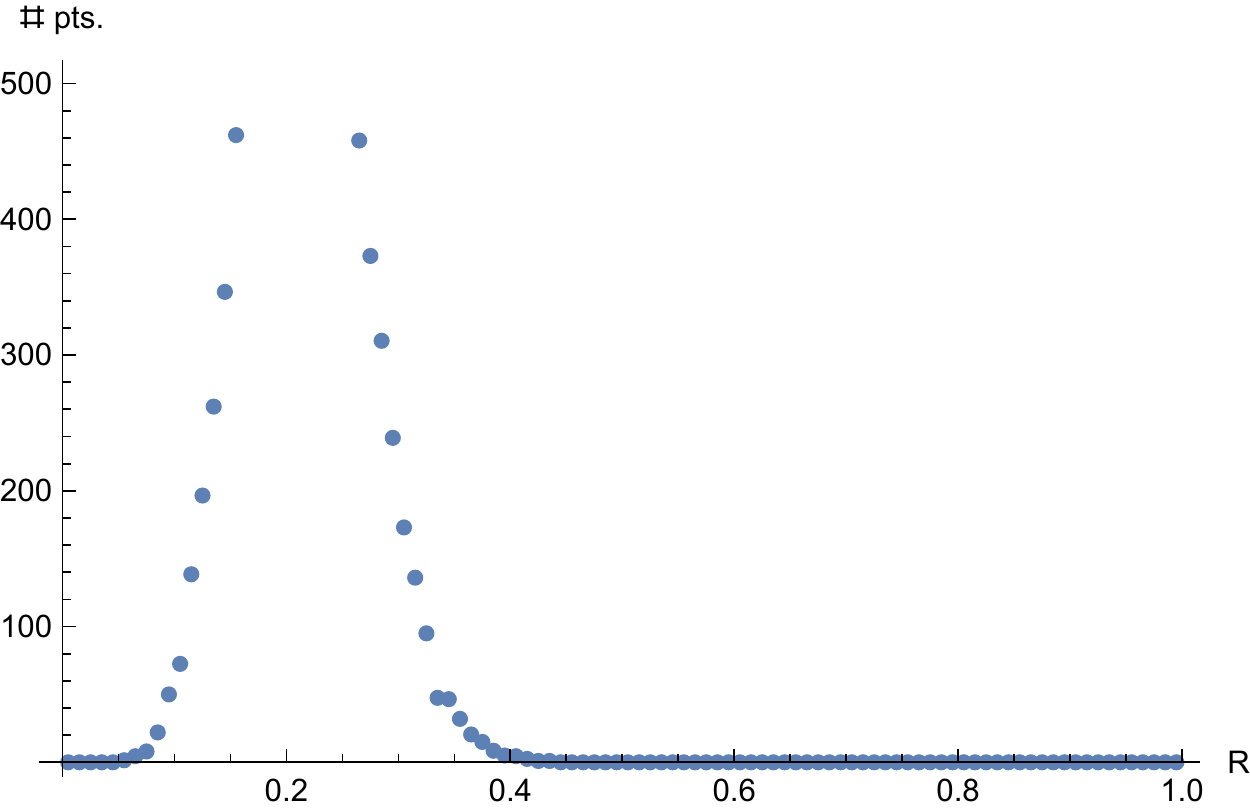}
\caption{\label{fig:PPTtwoqutrits}Distribution of sampled  positive-partial-transpose two-qutrit states over generalized Bloch radii $R$.}
\end{figure}

\begin{figure}
\includegraphics[scale=0.65]{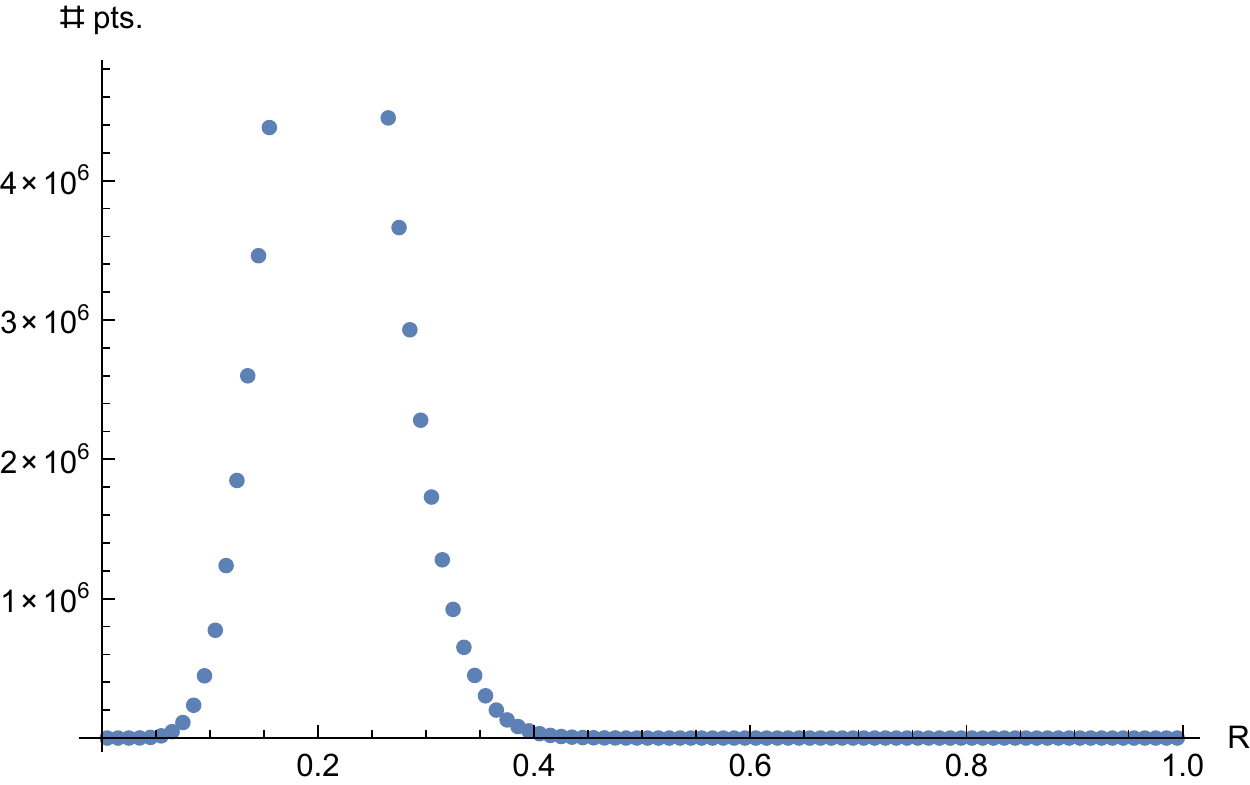}
\caption{\label{fig:Alltwoqutrits}Distribution of sampled  two-qutrit states over generalized Bloch radii $R$.}
\end{figure}

The resulting two-qutrit plot (Fig.~\ref{fig:TwoQutritsprob}) appears to be not inconsistent with
a hypothesis of constancy of PPT-probabilities along the generalized Bloch radius $R$.
This figure had been obtained by taking the ratio of the Fig.~\ref{fig:PPTtwoqutrits}  histogram to 
(the similarly shaped, again)  Fig.~\ref{fig:Alltwoqutrits} histogram. (No density matrices with $R>\frac{29}{50}$ were randomly generated, reflecting the relative rarity of states in this domain. The {\it zero} probabilities 
appearing near $R=0$ should not be troubling, since presumably the estimated $R$-invariant probability is so small [0.000101481] that--given the corresponding sample sizes--zero outcomes are, in fact, the most likely ones here.)
\subsection{Qubit-{\it qudit} analysis}
Continuing along such extended lines, we generated, randomly with respect to Hilbert-Schmidt measure, 348,500,000  $8 \times 8$ density matrices, analyzing them as
qubit-{\it qudit} ($2 \times 4$) systems. Of them, 450,386 had PPT's, leading to an associated PPT-probability of 0.0012923558. Again, as Milz and Strunz 
specifically discerned \cite[Fig. 5]{milzstrunz}, the plot
of probabilities over the qubit (three-dimensional-based) Bloch radius ($r$) had a very flat/invariant profile. 

\begin{figure}
    \includegraphics[scale=0.65]{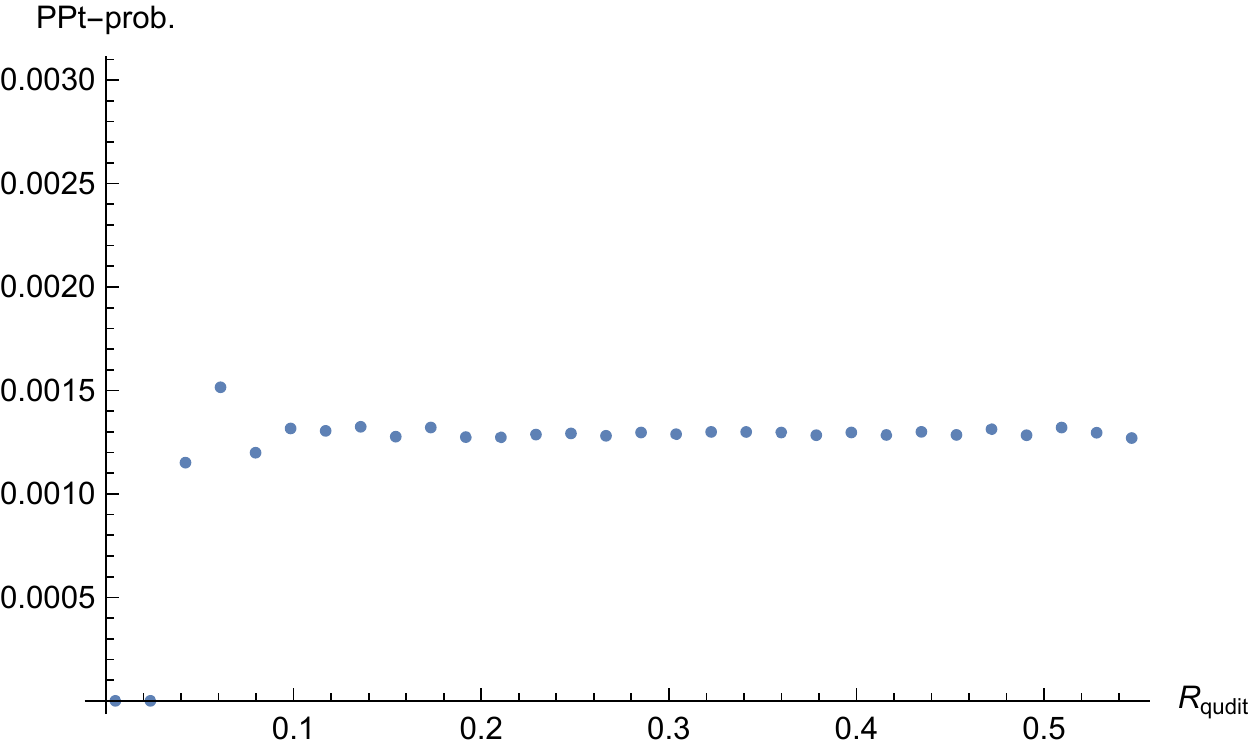}
    \caption{Plot of qubit-qudit ($2 \times 4$) PPT-probabilities over the length $R_{qudit}$ of the associated fifteen-dimensional Bloch vector}
    \label{fig:PPTqubitqudit}
\end{figure}

In Fig.~\ref{fig:PPTqubitqudit}, we now show the counterpart plot for the {\it qudit} (fifteen-dimensional-based) generalized Bloch radius
$R_{qudit}$. Again, consistently with our general findings here, that plot is similarly
flat.
\section{Higher-order (cubic) Casimir invariants}
Viewing the (generalized) Bloch radii in terms of {\it quadratic Casimir invariants} \cite{PhysRevA.68.062322,BoyaDixit}, perhaps it might be insightful
to employ the squares of the radii (that is, $r^2, R^2,\ldots$) as prinicipal variables themselves, rather than simply $r,R, \ldots$
Further, the possibility that invariance of separability (PPT-)probabilities continues to hold with respect to other (non-quadratic, cubic, \ldots) Casimir invariants
seems a topic well worthy of investigation, that we now pursue. 
\subsection{Second qubit-qutrit analysis} \label{2x3B}
\begin{figure}
    \includegraphics[scale=0.65]{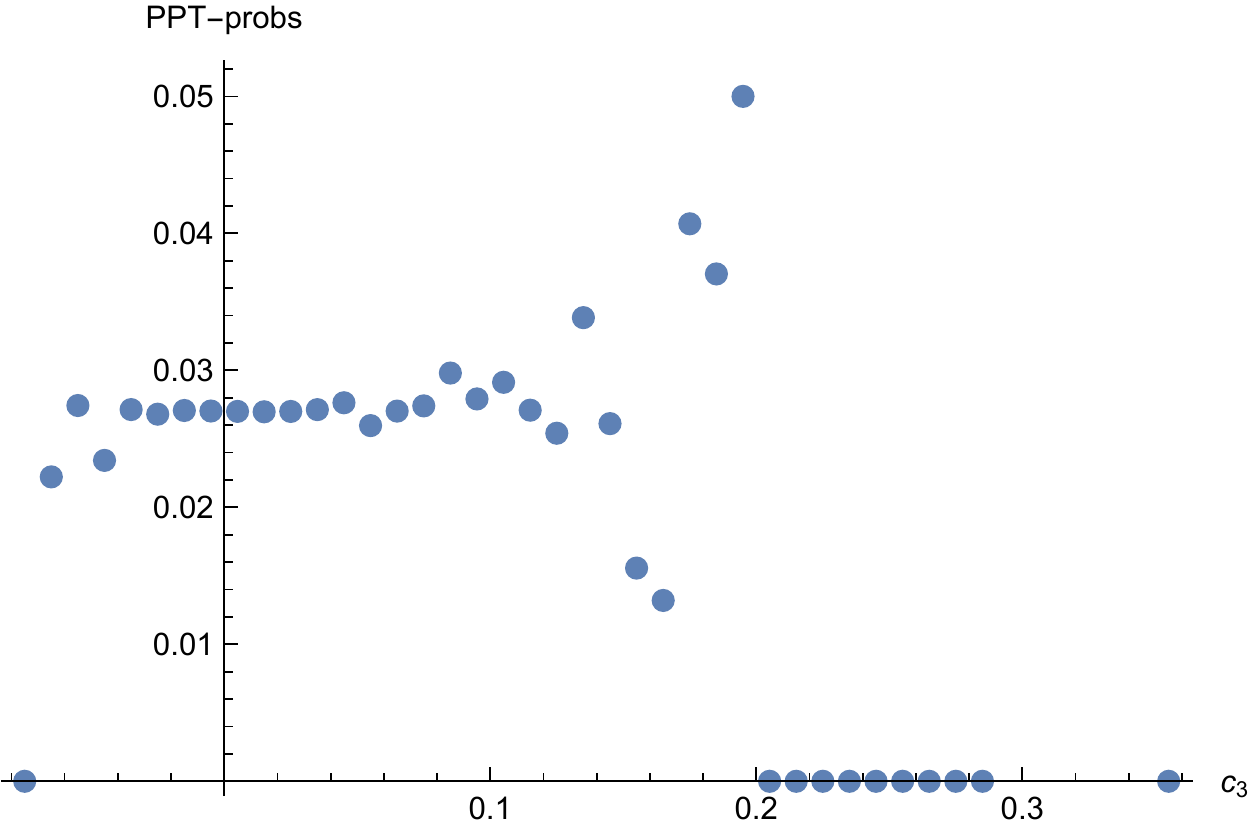}
    \caption{Qubit-qutrit separability probabilities as a function of the {\it cubic} Casimir invariant $c_3$}
    \label{fig:CasimirQutritprobs}
\end{figure}

We undertook a qubit-qutrit analysis parallel to that reported above (employing again, as in sec.~\ref{2x3A}, a  [new] sample of one hundred million 
random density matrices, of which 2,701,081 were separable.). But rather than
plotting (as in Fig.~\ref{fig:Qutritprobs}) the separability probabilities as a function of the qutrit Bloch radius $R$ (the square root of the corresponding {\it quadratic} 
Casimir invariant $c_2$), we utilized the corresponding {\it cubic} Casimir invariant $c_3$ \cite[eq. (35)]{PhysRevA.68.062322} \cite[eq. (11)]{Goyal} 
\cite[eq. (20)]{Gerdt2011}, 
\begin{equation} \label{cubicinvariant}
c_3 = \overrightarrow{n}* \overrightarrow{n} \cdot \overrightarrow{n},  
\end{equation}
where $\overrightarrow{n}$ is the Bloch 8-vector (and 
$c_2 = \overrightarrow{n} \cdot  \overrightarrow{n}$ is the square of the Bloch radius $R$).
The resulting plot is Fig.~\ref{fig:CasimirQutritprobs}.
(Again, we apparently see noisy scatter in the 
extreme upper and lower tails, having relatively low numbers of sampled density matrices.)
Pooling these results with those in sec.~\ref{2x3A}, we obtain a Hilbert-Schmidt qubit-qutrit separability probability estimate of 0.027003355, centered in 
the $95\%$ confidence interval $\{0.0269809,0.0270258\}$.
\section{Polynomial invariants in two-qubit case}
Byrd and Khaneja \cite{PhysRevA.68.062322} had observed that the number of polynomial invariants \cite{Makhlin} under unitary 
transformations  is larger than the number of Casimir invariants, which are included as a subset. So, we might pose the further question
of whether uniformity of separability (PPT-)probabilities holds too for any of the larger set of polynomial invariants.

Khvedelidze and Rogojin have listed (up to the fourth order) an ``integrity basis of SU(2) $\otimes$ SU(2) invariants in the enveloping algebra $\mathfrak{U(su(n))}$''. They, first, give three invariants of the {\it second} degree \cite[eq. (34)]{Khvedelidze2015} 
(also \cite[eq. (29)]{Gerdt2010}). The first two ($C^{(200)}, C^{(020)}$, in the notation of Quesne that they adopt), of the three,  are simply equivalent to the squares of the Bloch radii ($r_A^2, r_B^2$)--that is, the quadratic Casimir invariants. 

\begin{figure}
    \centering
    \includegraphics[scale=0.65]{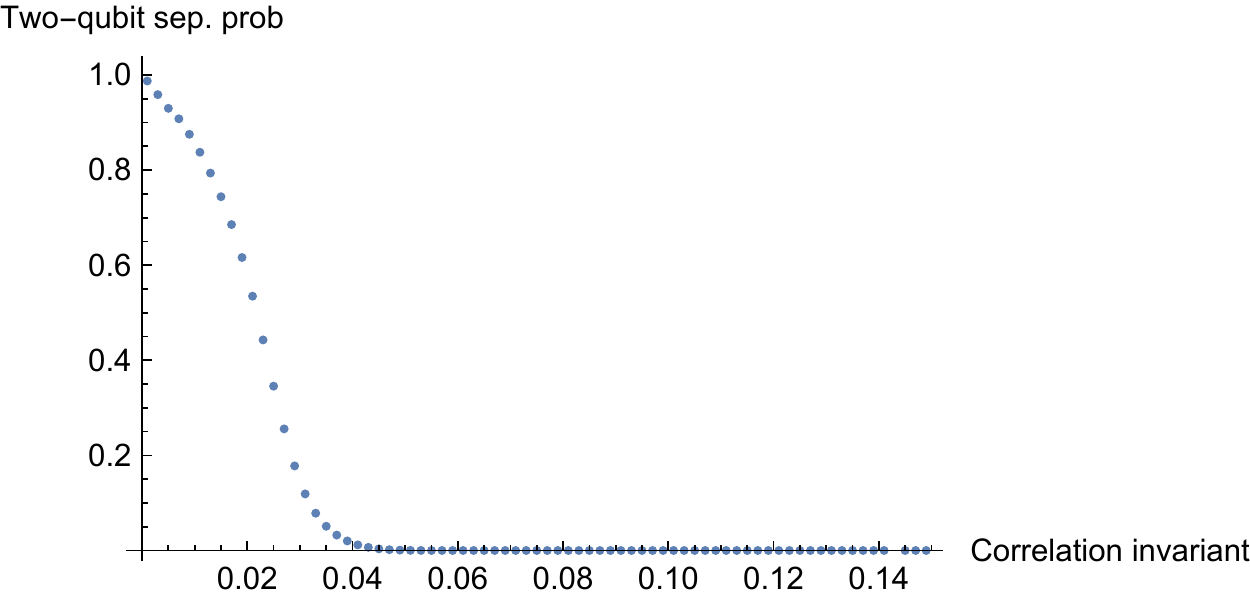}
    \caption{Two-qubit (clearly, now non-uniform) separability probabilities as a function of the second-degree correlation polynomial invariant $C^{(002)}$ given in eq. (\ref{correlationinvariant}).}
    \label{fig:CorrelationInvariant}
\end{figure}

So, we presumably know by the analyses of Milz and 
Strunz \cite{milzstrunz}, and the supporting evidence in \cite[Fig. 10]{Repulsion} that the Hilbert-Schmidt two-qubit separability probability is uniformly distributed at apparently $\frac{8}{33}$  over these two second-degree invariants. We are now interested in whether the separability probability is also uniform over the third of their (now, {\it non-local}) second-degree
polynomial invariants, namely
\begin{equation} \label{correlationinvariant}
C^{(002)} =c_{ij} c_{ij} = \Sigma_{i=1,j=1}^{i=3,j=3} c_{ij}.    
\end{equation}
The $c_{ij}$'s are the entries of the $3 \times 3$ ``correlation matrix'' in the well-known Fano decomposition
of a two-qubit state \cite[eq. (29)]{Khvedelidze2015} (they ``contain information on interactions between parts of a composite system'' 
\cite{Khvedelidze2015}). We have performed an analysis based on twenty million $4 \times 4$ density matrices,
randomly drawn with respect to Hilbert-Schmidt measure--with 4,843,346 of them being separable, yielding a separability probability
estimate of 0.2421673.	
The corresponding plot is Fig.~\ref{fig:CorrelationInvariant} (cf. \cite[Fig. 52]{Repulsion}).

Obviously, this plot constitutes, in general, compelling evidence against the invariance of
separability probabilities over (non-local) polynomial invariants, less specific than the partial/local Casimir ones.
Of course, it would be possible to similarly analyze the other (third- and fourth-degree, non-Casimir) invariant polynomials \cite[eqs. (35)-(39)]{Khvedelidze2015}, 
but we have no particular expectations that any single one might lead to uniformity of separability probabilities.
\section{Qubit-qutrit analyses with random induced measure}
One might additionally investigate--motivated by results of other recent studies \cite{LatestCollaboration,HyperDiff}--the issue of whether invariances 
such as those apparently observed above, continue to hold when, more generally, random induced measures \cite{Induced}, other than the specific (symmetric) Hilbert-Schmidt form of such measures are imposed.  
(Also, analyses in the real and quaternionic (cf. \cite{BG,FeiJoynt}) domains might be conducted.)

Let us now perform a {\it third} qubit-qutrit analysis, after those in secs.~\ref{2x3A} and \ref{2x3B}--in which we found evidence for the invariance of 
separability probabilities over both the generalized (qutrit) Bloch radius $R= \sqrt{c_2}$ and the cubic Casimir invariant $c_3$.
Those analyses were conducted using Hilbert-Schmidt measure, implicitly the {\it symmetric} instance, with
an ancillary Hilbert space of dimension six, that is, $N=K=6$, of more generally random induced measure \cite{Induced}. We, thus, modify the analyses by employing
an ancillary space of, we choose, dimension nine, $K=9$. 

\begin{figure}
    \centering
    \includegraphics[scale=0.65]{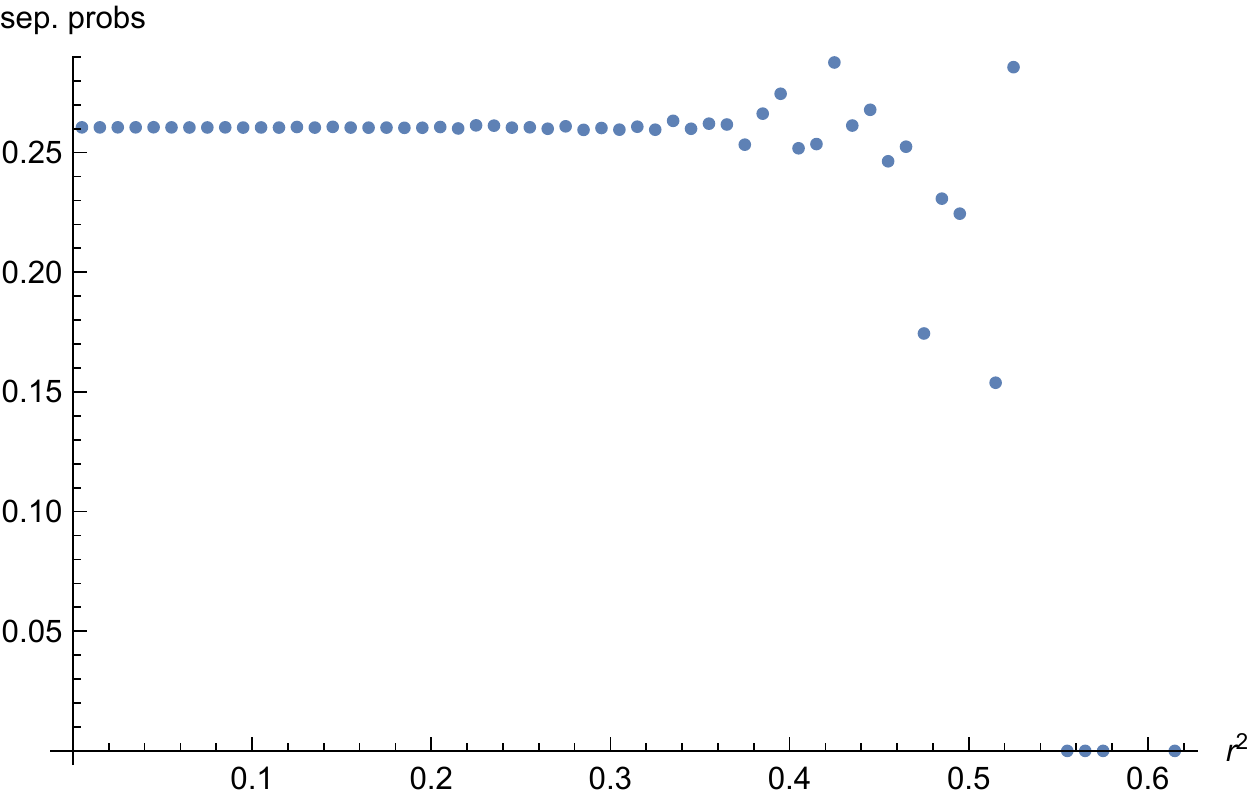}
    \caption{Random induced measure (N=6, K=9) qubit-qutrit separability probability estimates over the corresponding quadratic Casimir invariant, that is the 
    square ($r^2$) of the qubit Bloch radius.}
    \label{fig:Ancillaryqubit}
\end{figure}

\begin{figure}
    \centering
    \includegraphics[scale=0.65]{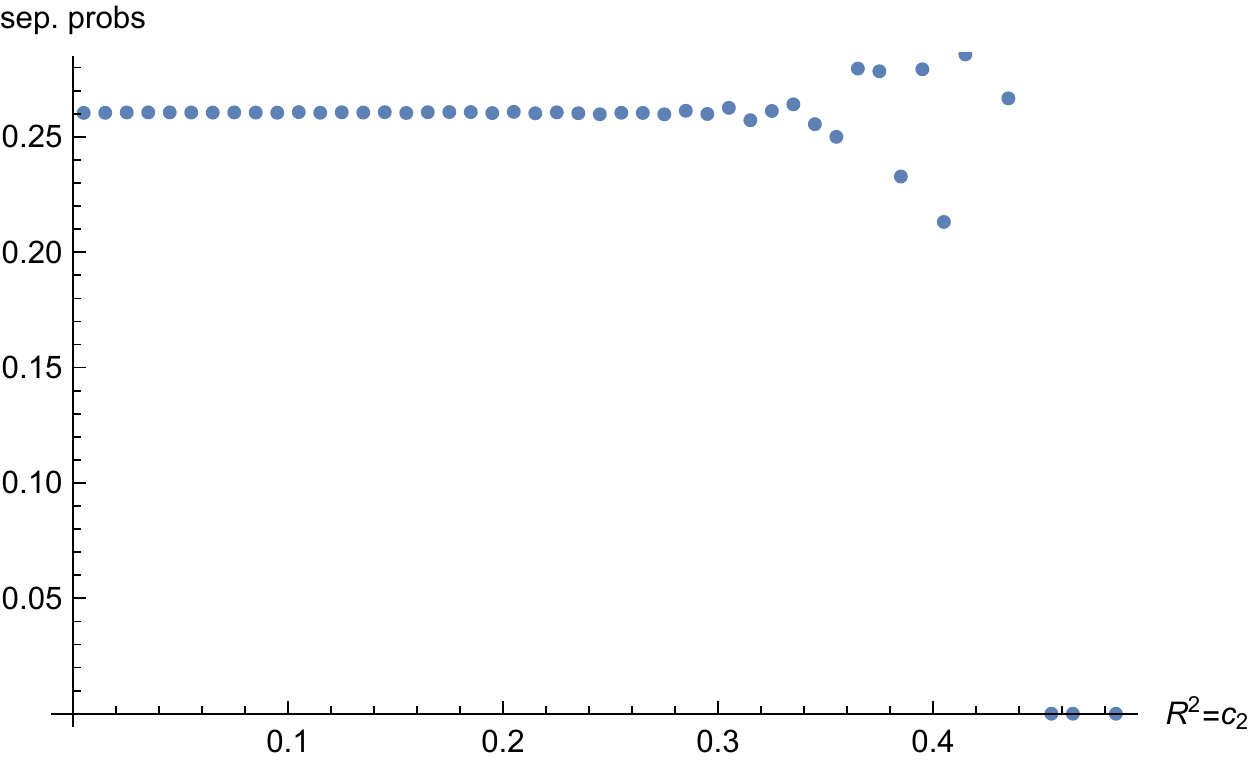}
    \caption{Random induced measure (N=6, K=9) qubit-qutrit separability probability estimates over the corresponding quadratic Casimir invariant, that is the 
    square ($R^2=c_2$) of the qutrit generalized Bloch radius.}
    \label{fig:AncillaryR2}
\end{figure}

\begin{figure}
    \centering
    \includegraphics[scale=0.65]{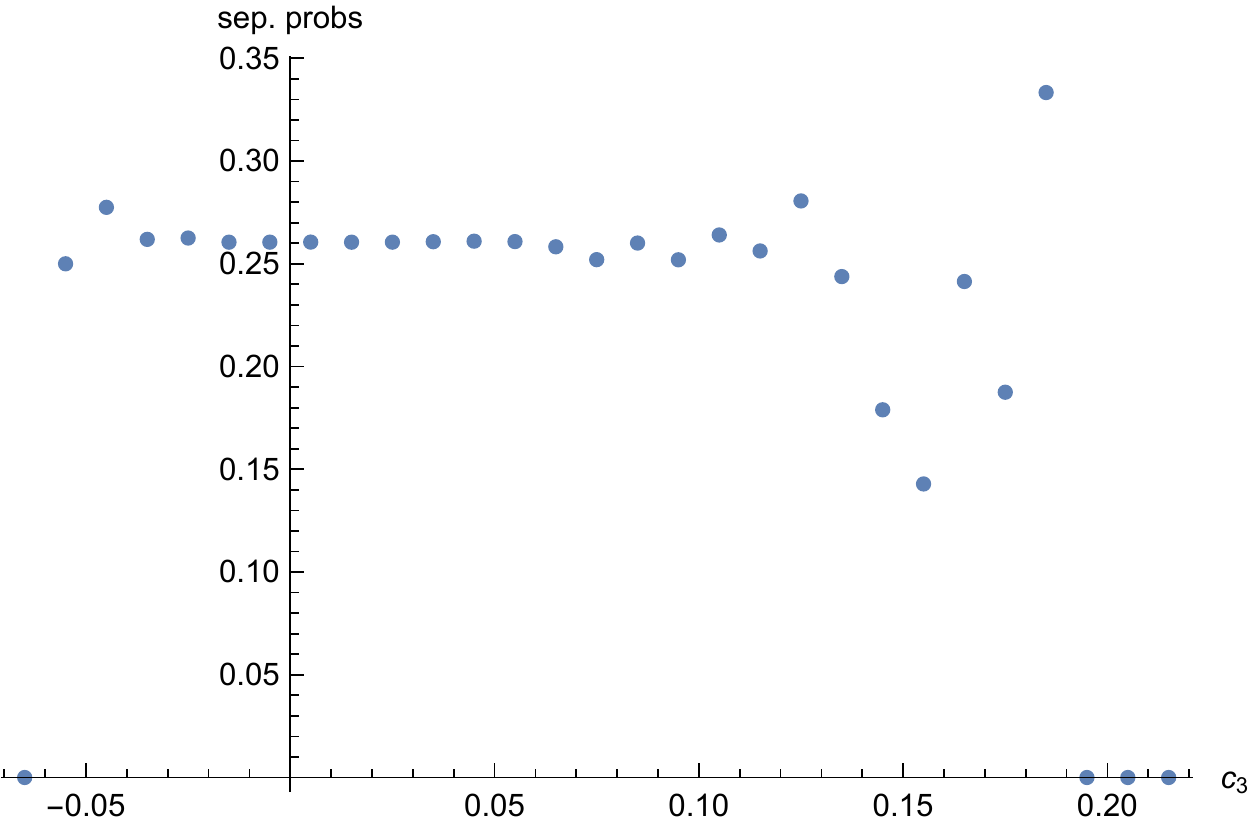}
    \caption{Random induced measure (N=6, K=9) qubit-qutrit separability probability estimates over the qutrit cubic Casimir invariant $c_3$, given
    in eq. (\ref{cubicinvariant}).}
    \label{fig:Ancillaryc3}
\end{figure}

1,097 million $6 \times 6$ density matrices were generated with respect to the corresponding measure (following the prescription in \cite{generating}).
285,823,317 of these were separable, yielding a separability probability estimate of 0.260549969917958. (The associated $95\%$ confidence interval was $\{0.260524,0.260576\}$.) For each density matrix, we recorded and binned the values of the {\it three} variables--$r^2$, $R^2=c_2$ and
$c_3$. In Figs.~\ref{fig:Ancillaryqubit} and \ref{fig:AncillaryR2} and \ref{fig:Ancillaryc3}, we plot the separability probability estimates as functions of these three variables.
These three plots--in particular, the first two--strongly indicate invariance of separability probability estimates over the corresponding Casimir invariants.

\section{Concluding remarks}
In the case of $8 \times 8$ (qubit-qudit) density matrices, there are {\it three} independent Casimir
invariants for the qudit subsystem  \cite[eqs. (46)-(48)]{Gerdt2010} \cite[eqs. (28)-(30)]{GKP2}, with the PPT-probabilities conjecturally holding
constant over {\it each} of the three invariants. (We already have acquired evidence [Fig.~\ref{fig:PPTqubitqudit}] as to apparent invariance
over $R_{qudit} =\sqrt{c_2}$.)

Let us note the existence of but only a limited body of formally rigorous results (theorems, lemmas,\ldots) pertaining
to properties of Hilbert-Schmidt (finite-dimensional) separability probabilities 
\cite{sbz,LatestCollaboration2} (cf. \cite{aubrun2}). Proofs are certainly still lacking for the interesting generalized Bloch radii/Casimir invariants conjectures made above, as well as those in a number of other recent reports \cite{slater833,MomentBased,slaterJModPhys,FeiJoynt,
LatestCollaboration,HyperDiff,Repulsion}, and the infinite summation formula 
given above ((\ref{Hou1})-(\ref{Hou3})).

On the other hand, clear evidence has been provided \cite[Fig. 31]{Repulsion} that the apparent $r$-invariance phenomenon revealed by the work of Milz and Strunz 
\cite{milzstrunz} and supported above and in \cite{Repulsion} does {\it not} continue to hold if one employs, rather than Hilbert-Schmidt measure, its Bures (minimal monotone) \cite{szBures} counterpart.

Let us also indicate the interesting paper of Altafini, entitled "Tensor of coherences parametrization of multiqubit density operators for entanglement characterization"  \cite{Altafini}. In 
it, he applies the term "partial quadratic Casimir invariant" in relation to {\it reduced} density matrices. He notes that a quadratic Casimir invariant can be regarded as the specific form ($q=2$) of Tsallis entropy. Further, he remarks that "partial transposition is a linear norm preserving operation: 
$\mbox{tr}(\rho^2)=\mbox{tr}((\rho^{T_1})^2) =\mbox{tr}((\rho^{T_2})^2)$. Hence entanglement violating PPT does not modify the quadratic Casimir invariants of the density and the necessary [separability] conditions [$\mbox{tr}(\rho_A^2) \geq \mbox{tr}(\rho^2), \mbox{tr}(\rho_B^2) \geq \mbox{tr}(\rho^2)$] are insensible to it".

The space  of two-qubit density matrices has been explicitly defined in terms of polynomial inequalities in the Casimir operators of the enveloping algebra of the 
$SU(4)$ group  \cite{GKP2}. Additionally, the Peres-Horodecki separability conditions have been given in the form of polynomial inequalities in three $SU(4)$ Casimir invariants and two $SU(2) \otimes SU(2)$ scalars; namely determinants of the so-called {\it correlation} and {\it Schlienz-Mahler entanglement} matrices.

\bibliography{Constancy8}

\begin{thebibliography}{42}%
\makeatletter
\providecommand \@ifxundefined [1]{%
 \@ifx{#1\undefined}
}%
\providecommand \@ifnum [1]{%
 \ifnum #1\expandafter \@firstoftwo
 \else \expandafter \@secondoftwo
 \fi
}%
\providecommand \@ifx [1]{%
 \ifx #1\expandafter \@firstoftwo
 \else \expandafter \@secondoftwo
 \fi
}%
\providecommand \natexlab [1]{#1}%
\providecommand \enquote  [1]{``#1''}%
\providecommand \bibnamefont  [1]{#1}%
\providecommand \bibfnamefont [1]{#1}%
\providecommand \citenamefont [1]{#1}%
\providecommand \href@noop [0]{\@secondoftwo}%
\providecommand \href [0]{\begingroup \@sanitize@url \@href}%
\providecommand \@href[1]{\@@startlink{#1}\@@href}%
\providecommand \@@href[1]{\endgroup#1\@@endlink}%
\providecommand \@sanitize@url [0]{\catcode `\\12\catcode `\$12\catcode
  `\&12\catcode `\#12\catcode `\^12\catcode `\_12\catcode `\%12\relax}%
\providecommand \@@startlink[1]{}%
\providecommand \@@endlink[0]{}%
\providecommand \url  [0]{\begingroup\@sanitize@url \@url }%
\providecommand \@url [1]{\endgroup\@href {#1}{\urlprefix }}%
\providecommand \urlprefix  [0]{URL }%
\providecommand \Eprint [0]{\href }%
\providecommand \doibase [0]{http://dx.doi.org/}%
\providecommand \selectlanguage [0]{\@gobble}%
\providecommand \bibinfo  [0]{\@secondoftwo}%
\providecommand \bibfield  [0]{\@secondoftwo}%
\providecommand \translation [1]{[#1]}%
\providecommand \BibitemOpen [0]{}%
\providecommand \bibitemStop [0]{}%
\providecommand \bibitemNoStop [0]{.\EOS\space}%
\providecommand \EOS [0]{\spacefactor3000\relax}%
\providecommand \BibitemShut  [1]{\csname bibitem#1\endcsname}%
\let\auto@bib@innerbib\@empty
\bibitem [{\citenamefont {Gamel}(2016)}]{Gamel}%
  \BibitemOpen
  \bibfield  {author} {\bibinfo {author} {\bibfnamefont {O.}~\bibnamefont
  {Gamel}},\ }\href@noop {} {\bibfield  {journal} {\bibinfo  {journal} {arXiv
  preprint arXiv:1602.01548}\ } (\bibinfo {year} {2016})}\BibitemShut {NoStop}%
\bibitem [{\citenamefont {Bengtsson}\ and\ \citenamefont
  {{\.Z}yczkowski}(2006)}]{ingemarkarol}%
  \BibitemOpen
  \bibfield  {author} {\bibinfo {author} {\bibfnamefont {I.}~\bibnamefont
  {Bengtsson}}\ and\ \bibinfo {author} {\bibfnamefont {K.}~\bibnamefont
  {{\.Z}yczkowski}},\ }\href@noop {} {\emph {\bibinfo {title} {Geometry of
  Quantum States}}}\ (\bibinfo  {publisher} {Cambridge},\ \bibinfo {address}
  {Cambridge},\ \bibinfo {year} {2006})\BibitemShut {NoStop}%
\bibitem [{\citenamefont {Petz}\ and\ \citenamefont
  {Sud{\'a}r}(1996)}]{petzsudar}%
  \BibitemOpen
  \bibfield  {author} {\bibinfo {author} {\bibfnamefont {D.}~\bibnamefont
  {Petz}}\ and\ \bibinfo {author} {\bibfnamefont {C.}~\bibnamefont
  {Sud{\'a}r}},\ }\href@noop {} {\bibfield  {journal} {\bibinfo  {journal} {J.
  Math. Phys.}\ }\textbf {\bibinfo {volume} {37}},\ \bibinfo {pages} {2662}
  (\bibinfo {year} {1996})}\BibitemShut {NoStop}%
\bibitem [{\citenamefont {{\.Z}yczkowski}\ \emph {et~al.}(1998)\citenamefont
  {{\.Z}yczkowski}, \citenamefont {Horodecki}, \citenamefont {Sanpera},\ and\
  \citenamefont {Lewenstein}}]{ZHSL}%
  \BibitemOpen
  \bibfield  {author} {\bibinfo {author} {\bibfnamefont {K.}~\bibnamefont
  {{\.Z}yczkowski}}, \bibinfo {author} {\bibfnamefont {P.}~\bibnamefont
  {Horodecki}}, \bibinfo {author} {\bibfnamefont {A.}~\bibnamefont {Sanpera}},
  \ and\ \bibinfo {author} {\bibfnamefont {M.}~\bibnamefont {Lewenstein}},\
  }\href@noop {} {\bibfield  {journal} {\bibinfo  {journal} {Phys. Rev. A}\
  }\textbf {\bibinfo {volume} {58}},\ \bibinfo {pages} {883} (\bibinfo {year}
  {1998})}\BibitemShut {NoStop}%
\bibitem [{\citenamefont {Szarek}\ \emph {et~al.}(2006)\citenamefont {Szarek},
  \citenamefont {Bengtsson},\ and\ \citenamefont {{\.Z}yczkowski}}]{sbz}%
  \BibitemOpen
  \bibfield  {author} {\bibinfo {author} {\bibfnamefont {S.}~\bibnamefont
  {Szarek}}, \bibinfo {author} {\bibfnamefont {I.}~\bibnamefont {Bengtsson}}, \
  and\ \bibinfo {author} {\bibfnamefont {K.}~\bibnamefont {{\.Z}yczkowski}},\
  }\href@noop {} {\bibfield  {journal} {\bibinfo  {journal} {J. Phys. A}\
  }\textbf {\bibinfo {volume} {39}},\ \bibinfo {pages} {L119} (\bibinfo {year}
  {2006})}\BibitemShut {NoStop}%
\bibitem [{\citenamefont {Gerdt}\ \emph
  {et~al.}(2011{\natexlab{a}})\citenamefont {Gerdt}, \citenamefont {Mladenov},
  \citenamefont {Palii},\ and\ \citenamefont {Khvedelidze}}]{Gerdt2011}%
  \BibitemOpen
  \bibfield  {author} {\bibinfo {author} {\bibfnamefont {V.}~\bibnamefont
  {Gerdt}}, \bibinfo {author} {\bibfnamefont {D.}~\bibnamefont {Mladenov}},
  \bibinfo {author} {\bibfnamefont {Y.}~\bibnamefont {Palii}}, \ and\ \bibinfo
  {author} {\bibfnamefont {A.}~\bibnamefont {Khvedelidze}},\ }\href {\doibase
  10.1007/s10958-011-0619-9} {\bibfield  {journal} {\bibinfo  {journal}
  {Journal of Mathematical Sciences}\ }\textbf {\bibinfo {volume} {179}},\
  \bibinfo {pages} {690} (\bibinfo {year} {2011}{\natexlab{a}})}\BibitemShut
  {NoStop}%
\bibitem [{\citenamefont {Goyal}\ \emph {et~al.}(2011)\citenamefont {Goyal},
  \citenamefont {Simon}, \citenamefont {Singh},\ and\ \citenamefont
  {Simon}}]{Goyal}%
  \BibitemOpen
  \bibfield  {author} {\bibinfo {author} {\bibfnamefont {S.~K.}\ \bibnamefont
  {Goyal}}, \bibinfo {author} {\bibfnamefont {B.~N.}\ \bibnamefont {Simon}},
  \bibinfo {author} {\bibfnamefont {R.}~\bibnamefont {Singh}}, \ and\ \bibinfo
  {author} {\bibfnamefont {S.}~\bibnamefont {Simon}},\ }\href@noop {}
  {\bibfield  {journal} {\bibinfo  {journal} {arXiv preprint arXiv:1111.4427}\
  } (\bibinfo {year} {2011})}\BibitemShut {NoStop}%
\bibitem [{\citenamefont {Kimura}(2003)}]{Kimura}%
  \BibitemOpen
  \bibfield  {author} {\bibinfo {author} {\bibfnamefont {G.}~\bibnamefont
  {Kimura}},\ }\href@noop {} {\bibfield  {journal} {\bibinfo  {journal} {Phys.
  Lett. A}\ }\textbf {\bibinfo {volume} {314}},\ \bibinfo {pages} {339}
  (\bibinfo {year} {2003})}\BibitemShut {NoStop}%
\bibitem [{\citenamefont {Scutaru}(2005)}]{scutaru2}%
  \BibitemOpen
  \bibfield  {author} {\bibinfo {author} {\bibfnamefont {H.}~\bibnamefont
  {Scutaru}},\ }\href@noop {} {\bibfield  {journal} {\bibinfo  {journal} {Proc.
  Romanian Acad. Sci.}\ }\textbf {\bibinfo {volume} {6}},\ \bibinfo {pages}
  {000} (\bibinfo {year} {2005})}\BibitemShut {NoStop}%
\bibitem [{\citenamefont {Werner}(1989)}]{ClassicallyCorrelated}%
  \BibitemOpen
  \bibfield  {author} {\bibinfo {author} {\bibfnamefont {R.~F.}\ \bibnamefont
  {Werner}},\ }\href {\doibase 10.1103/PhysRevA.40.4277} {\bibfield  {journal}
  {\bibinfo  {journal} {Phys. Rev. A}\ }\textbf {\bibinfo {volume} {40}},\
  \bibinfo {pages} {4277} (\bibinfo {year} {1989})}\BibitemShut {NoStop}%
\bibitem [{\citenamefont {Slater}(2007)}]{slater833}%
  \BibitemOpen
  \bibfield  {author} {\bibinfo {author} {\bibfnamefont {P.~B.}\ \bibnamefont
  {Slater}},\ }\href@noop {} {\bibfield  {journal} {\bibinfo  {journal} {J.
  Phys. A}\ }\textbf {\bibinfo {volume} {40}},\ \bibinfo {pages} {14279}
  (\bibinfo {year} {2007})}\BibitemShut {NoStop}%
\bibitem [{\citenamefont {Slater}\ and\ \citenamefont
  {Dunkl}(2012)}]{MomentBased}%
  \BibitemOpen
  \bibfield  {author} {\bibinfo {author} {\bibfnamefont {P.~B.}\ \bibnamefont
  {Slater}}\ and\ \bibinfo {author} {\bibfnamefont {C.~F.}\ \bibnamefont
  {Dunkl}},\ }\href@noop {} {\bibfield  {journal} {\bibinfo  {journal} {J.
  Phys. A}\ }\textbf {\bibinfo {volume} {45}},\ \bibinfo {pages} {095305}
  (\bibinfo {year} {2012})}\BibitemShut {NoStop}%
\bibitem [{\citenamefont {Slater}(2013)}]{slaterJModPhys}%
  \BibitemOpen
  \bibfield  {author} {\bibinfo {author} {\bibfnamefont {P.~B.}\ \bibnamefont
  {Slater}},\ }\href@noop {} {\bibfield  {journal} {\bibinfo  {journal} {J.
  Phys. A}\ }\textbf {\bibinfo {volume} {46}},\ \bibinfo {pages} {445302}
  (\bibinfo {year} {2013})}\BibitemShut {NoStop}%
\bibitem [{\citenamefont {Fei}\ and\ \citenamefont {Joynt}(2014)}]{FeiJoynt}%
  \BibitemOpen
  \bibfield  {author} {\bibinfo {author} {\bibfnamefont {J.}~\bibnamefont
  {Fei}}\ and\ \bibinfo {author} {\bibfnamefont {R.}~\bibnamefont {Joynt}},\
  }\href@noop {} {\bibfield  {journal} {\bibinfo  {journal} {arXiv preprint
  arXiv:1409.1993}\ } (\bibinfo {year} {2014})}\BibitemShut {NoStop}%
\bibitem [{\citenamefont {{\.Z}yczkowski}\ and\ \citenamefont
  {Sommers}(2003)}]{szHS}%
  \BibitemOpen
  \bibfield  {author} {\bibinfo {author} {\bibfnamefont {K.}~\bibnamefont
  {{\.Z}yczkowski}}\ and\ \bibinfo {author} {\bibfnamefont {H.-J.}\
  \bibnamefont {Sommers}},\ }\href@noop {} {\bibfield  {journal} {\bibinfo
  {journal} {J. Phys. A}\ }\textbf {\bibinfo {volume} {36}},\ \bibinfo {pages}
  {10115} (\bibinfo {year} {2003})}\BibitemShut {NoStop}%
\bibitem [{\citenamefont {{\.Z}yczkowski}\ and\ \citenamefont
  {Sommers}(2001)}]{Induced}%
  \BibitemOpen
  \bibfield  {author} {\bibinfo {author} {\bibfnamefont {K.}~\bibnamefont
  {{\.Z}yczkowski}}\ and\ \bibinfo {author} {\bibfnamefont {H.-J.}\
  \bibnamefont {Sommers}},\ }\href@noop {} {\bibfield  {journal} {\bibinfo
  {journal} {J. Phys. A}\ }\textbf {\bibinfo {volume} {A34}},\ \bibinfo {pages}
  {7111} (\bibinfo {year} {2001})}\BibitemShut {NoStop}%
\bibitem [{\citenamefont {Zeilberger}(1990)}]{doron}%
  \BibitemOpen
  \bibfield  {author} {\bibinfo {author} {\bibfnamefont {D.}~\bibnamefont
  {Zeilberger}},\ }\href@noop {} {\bibfield  {journal} {\bibinfo  {journal}
  {Discr. Math.}\ }\textbf {\bibinfo {volume} {80}},\ \bibinfo {pages} {207}
  (\bibinfo {year} {1990})}\BibitemShut {NoStop}%
\bibitem [{\citenamefont {Dumitriu}\ and\ \citenamefont
  {Edelman}(2002)}]{MatrixModels}%
  \BibitemOpen
  \bibfield  {author} {\bibinfo {author} {\bibfnamefont {I.}~\bibnamefont
  {Dumitriu}}\ and\ \bibinfo {author} {\bibfnamefont {A.}~\bibnamefont
  {Edelman}},\ }\href@noop {} {\bibfield  {journal} {\bibinfo  {journal} {J.
  Math. Phys.}\ }\textbf {\bibinfo {volume} {43}},\ \bibinfo {pages} {5830}
  (\bibinfo {year} {2002})}\BibitemShut {NoStop}%
\bibitem [{\citenamefont {Milz}\ and\ \citenamefont
  {Strunz}(2015)}]{milzstrunz}%
  \BibitemOpen
  \bibfield  {author} {\bibinfo {author} {\bibfnamefont {S.}~\bibnamefont
  {Milz}}\ and\ \bibinfo {author} {\bibfnamefont {W.~T.}\ \bibnamefont
  {Strunz}},\ }\href@noop {} {\bibfield  {journal} {\bibinfo  {journal} {J.
  Phys. A}\ }\textbf {\bibinfo {volume} {48}},\ \bibinfo {pages} {035306}
  (\bibinfo {year} {2015})}\BibitemShut {NoStop}%
\bibitem [{\citenamefont {Slater}\ and\ \citenamefont
  {Dunkl}(2015)}]{LatestCollaboration}%
  \BibitemOpen
  \bibfield  {author} {\bibinfo {author} {\bibfnamefont {P.~B.}\ \bibnamefont
  {Slater}}\ and\ \bibinfo {author} {\bibfnamefont {C.~F.}\ \bibnamefont
  {Dunkl}},\ }\href@noop {} {\bibfield  {journal} {\bibinfo  {journal} {Adv.
  Math. Phys.}\ }\textbf {\bibinfo {volume} {2015}},\ \bibinfo {pages} {621353}
  (\bibinfo {year} {2015})}\BibitemShut {NoStop}%
\bibitem [{\citenamefont {Jevtic}\ \emph {et~al.}(2014)\citenamefont {Jevtic},
  \citenamefont {Pusey}, \citenamefont {Jennings},\ and\ \citenamefont
  {Rudolph}}]{Jevtic}%
  \BibitemOpen
  \bibfield  {author} {\bibinfo {author} {\bibfnamefont {S.}~\bibnamefont
  {Jevtic}}, \bibinfo {author} {\bibfnamefont {M.}~\bibnamefont {Pusey}},
  \bibinfo {author} {\bibfnamefont {D.}~\bibnamefont {Jennings}}, \ and\
  \bibinfo {author} {\bibfnamefont {T.}~\bibnamefont {Rudolph}},\ }\href@noop
  {} {\bibfield  {journal} {\bibinfo  {journal} {Phys. Rev. Lett.}\ }\textbf
  {\bibinfo {volume} {113}},\ \bibinfo {pages} {020402} (\bibinfo {year}
  {2014})}\BibitemShut {NoStop}%
\bibitem [{\citenamefont {Slater}(2015{\natexlab{a}})}]{Repulsion}%
  \BibitemOpen
  \bibfield  {author} {\bibinfo {author} {\bibfnamefont {P.~B.}\ \bibnamefont
  {Slater}},\ }\href@noop {} {\bibfield  {journal} {\bibinfo  {journal} {arXiv
  preprint arXiv:1506.08739}\ } (\bibinfo {year}
  {2015}{\natexlab{a}})}\BibitemShut {NoStop}%
\bibitem [{\citenamefont {Byrd}\ and\ \citenamefont
  {Khaneja}(2003)}]{PhysRevA.68.062322}%
  \BibitemOpen
  \bibfield  {author} {\bibinfo {author} {\bibfnamefont {M.~S.}\ \bibnamefont
  {Byrd}}\ and\ \bibinfo {author} {\bibfnamefont {N.}~\bibnamefont {Khaneja}},\
  }\href {\doibase 10.1103/PhysRevA.68.062322} {\bibfield  {journal} {\bibinfo
  {journal} {Phys. Rev. A}\ }\textbf {\bibinfo {volume} {68}},\ \bibinfo
  {pages} {062322} (\bibinfo {year} {2003})}\BibitemShut {NoStop}%
\bibitem [{\citenamefont {De~Vicente}(2007)}]{DeVicente}%
  \BibitemOpen
  \bibfield  {author} {\bibinfo {author} {\bibfnamefont {J.~I.}\ \bibnamefont
  {De~Vicente}},\ }\href {http://dl.acm.org/citation.cfm?id=2011734.2011739}
  {\bibfield  {journal} {\bibinfo  {journal} {Quantum Info. Comput.}\ }\textbf
  {\bibinfo {volume} {7}},\ \bibinfo {pages} {624} (\bibinfo {year}
  {2007})}\BibitemShut {NoStop}%
\bibitem [{\citenamefont {{\.Z}yczkowski}\ \emph {et~al.}(2011)\citenamefont
  {{\.Z}yczkowski}, \citenamefont {Penson}, \citenamefont {Nechita},\ and\
  \citenamefont {Collins}}]{generating}%
  \BibitemOpen
  \bibfield  {author} {\bibinfo {author} {\bibfnamefont {K.}~\bibnamefont
  {{\.Z}yczkowski}}, \bibinfo {author} {\bibfnamefont {K.~A.}\ \bibnamefont
  {Penson}}, \bibinfo {author} {\bibfnamefont {I.}~\bibnamefont {Nechita}}, \
  and\ \bibinfo {author} {\bibfnamefont {B.}~\bibnamefont {Collins}},\
  }\href@noop {} {\bibfield  {journal} {\bibinfo  {journal} {J. Math. Phys.}\
  }\textbf {\bibinfo {volume} {52}},\ \bibinfo {pages} {062201} (\bibinfo
  {year} {2011})}\BibitemShut {NoStop}%
\bibitem [{\citenamefont {Peres}(1996)}]{asher}%
  \BibitemOpen
  \bibfield  {author} {\bibinfo {author} {\bibfnamefont {A.}~\bibnamefont
  {Peres}},\ }\href@noop {} {\bibfield  {journal} {\bibinfo  {journal} {Phys.
  Rev. Lett.}\ }\textbf {\bibinfo {volume} {77}},\ \bibinfo {pages} {1413}
  (\bibinfo {year} {1996})}\BibitemShut {NoStop}%
\bibitem [{\citenamefont {Horodecki}\ \emph {et~al.}(1996)\citenamefont
  {Horodecki}, \citenamefont {Horodecki},\ and\ \citenamefont
  {Horodecki}}]{michal}%
  \BibitemOpen
  \bibfield  {author} {\bibinfo {author} {\bibfnamefont {M.}~\bibnamefont
  {Horodecki}}, \bibinfo {author} {\bibfnamefont {P.}~\bibnamefont
  {Horodecki}}, \ and\ \bibinfo {author} {\bibfnamefont {R.}~\bibnamefont
  {Horodecki}},\ }\href@noop {} {\bibfield  {journal} {\bibinfo  {journal}
  {Phys. Lett. A}\ }\textbf {\bibinfo {volume} {223}},\ \bibinfo {pages} {1}
  (\bibinfo {year} {1996})}\BibitemShut {NoStop}%
\bibitem [{\citenamefont {Krishnamoorthy}\ and\ \citenamefont
  {Peng}(2008)}]{KP}%
  \BibitemOpen
  \bibfield  {author} {\bibinfo {author} {\bibfnamefont {K.}~\bibnamefont
  {Krishnamoorthy}}\ and\ \bibinfo {author} {\bibfnamefont {J.}~\bibnamefont
  {Peng}},\ }\href@noop {} {\bibfield  {journal} {\bibinfo  {journal} {J. Appl.
  Statist. Sci.}\ }\textbf {\bibinfo {volume} {16}},\ \bibinfo {pages} {23}
  (\bibinfo {year} {2008})}\BibitemShut {NoStop}%
\bibitem [{\citenamefont {V{\'e}rtesi}\ and\ \citenamefont
  {Brunner}(2014)}]{VertesiBrunner}%
  \BibitemOpen
  \bibfield  {author} {\bibinfo {author} {\bibfnamefont {T.}~\bibnamefont
  {V{\'e}rtesi}}\ and\ \bibinfo {author} {\bibfnamefont {N.}~\bibnamefont
  {Brunner}},\ }\href@noop {} {\bibfield  {journal} {\bibinfo  {journal}
  {Nature communications}\ }\textbf {\bibinfo {volume} {5}},\ \bibinfo {pages}
  {5297} (\bibinfo {year} {2014})}\BibitemShut {NoStop}%
\bibitem [{\citenamefont {Jarvis}(2014)}]{Jarvis}%
  \BibitemOpen
  \bibfield  {author} {\bibinfo {author} {\bibfnamefont {P.~D.}\ \bibnamefont
  {Jarvis}},\ }\href {http://stacks.iop.org/1751-8121/47/i=21/a=215302}
  {\bibfield  {journal} {\bibinfo  {journal} {Journal of Physics A:
  Mathematical and Theoretical}\ }\textbf {\bibinfo {volume} {47}},\ \bibinfo
  {pages} {215302} (\bibinfo {year} {2014})}\BibitemShut {NoStop}%
\bibitem [{\citenamefont {Brown}\ \emph {et~al.}(2001)\citenamefont {Brown},
  \citenamefont {Cai},\ and\ \citenamefont {DasGupta}}]{Brown}%
  \BibitemOpen
  \bibfield  {author} {\bibinfo {author} {\bibfnamefont {L.~D.}\ \bibnamefont
  {Brown}}, \bibinfo {author} {\bibfnamefont {T.~T.}\ \bibnamefont {Cai}}, \
  and\ \bibinfo {author} {\bibfnamefont {A.}~\bibnamefont {DasGupta}},\
  }\href@noop {} {\bibfield  {journal} {\bibinfo  {journal} {Statistical
  science}\ }\textbf {\bibinfo {volume} {16}},\ \bibinfo {pages} {101}
  (\bibinfo {year} {2001})}\BibitemShut {NoStop}%
\bibitem [{\citenamefont {Boya}\ and\ \citenamefont {Dixit}(2008)}]{BoyaDixit}%
  \BibitemOpen
  \bibfield  {author} {\bibinfo {author} {\bibfnamefont {L.~J.}\ \bibnamefont
  {Boya}}\ and\ \bibinfo {author} {\bibfnamefont {K.}~\bibnamefont {Dixit}},\
  }\href@noop {} {\bibfield  {journal} {\bibinfo  {journal} {Phys. Rev. A}\
  }\textbf {\bibinfo {volume} {78}},\ \bibinfo {pages} {042108} (\bibinfo
  {year} {2008})}\BibitemShut {NoStop}%
\bibitem [{\citenamefont {Makhlin}(2002)}]{Makhlin}%
  \BibitemOpen
  \bibfield  {author} {\bibinfo {author} {\bibfnamefont {Y.}~\bibnamefont
  {Makhlin}},\ }\href {\doibase 10.1023/A:1022144002391} {\bibfield  {journal}
  {\bibinfo  {journal} {Quantum Information Processing}\ }\textbf {\bibinfo
  {volume} {1}},\ \bibinfo {pages} {243} (\bibinfo {year} {2002})}\BibitemShut
  {NoStop}%
\bibitem [{\citenamefont {Khvedelidze}\ and\ \citenamefont
  {Rogojin}(2015)}]{Khvedelidze2015}%
  \BibitemOpen
  \bibfield  {author} {\bibinfo {author} {\bibfnamefont {A.}~\bibnamefont
  {Khvedelidze}}\ and\ \bibinfo {author} {\bibfnamefont {I.}~\bibnamefont
  {Rogojin}},\ }\href@noop {} {\bibfield  {journal} {\bibinfo  {journal}
  {Journal of Mathematical Sciences}\ }\textbf {\bibinfo {volume} {209}},\
  \bibinfo {pages} {988} (\bibinfo {year} {2015})}\BibitemShut {NoStop}%
\bibitem [{\citenamefont {Gerdt}\ \emph {et~al.}(2010)\citenamefont {Gerdt},
  \citenamefont {Khvedelidze},\ and\ \citenamefont {Palii}}]{Gerdt2010}%
  \BibitemOpen
  \bibfield  {author} {\bibinfo {author} {\bibfnamefont {V.}~\bibnamefont
  {Gerdt}}, \bibinfo {author} {\bibfnamefont {A.}~\bibnamefont {Khvedelidze}},
  \ and\ \bibinfo {author} {\bibfnamefont {Y.}~\bibnamefont {Palii}},\
  }\href@noop {} {\bibfield  {journal} {\bibinfo  {journal} {Journal of
  Mathematical Sciences}\ }\textbf {\bibinfo {volume} {168}},\ \bibinfo {pages}
  {368} (\bibinfo {year} {2010})}\BibitemShut {NoStop}%
\bibitem [{\citenamefont {Slater}(2015{\natexlab{b}})}]{HyperDiff}%
  \BibitemOpen
  \bibfield  {author} {\bibinfo {author} {\bibfnamefont {P.~B.}\ \bibnamefont
  {Slater}},\ }\href@noop {} {\bibfield  {journal} {\bibinfo  {journal} {arXiv
  preprint arXiv:1504.04555}\ } (\bibinfo {year}
  {2015}{\natexlab{b}})}\BibitemShut {NoStop}%
\bibitem [{\citenamefont {Brody}\ and\ \citenamefont {Graefe}(2011)}]{BG}%
  \BibitemOpen
  \bibfield  {author} {\bibinfo {author} {\bibfnamefont {D.~C.}\ \bibnamefont
  {Brody}}\ and\ \bibinfo {author} {\bibfnamefont {E.-M.}\ \bibnamefont
  {Graefe}},\ }\href@noop {} {\bibfield  {journal} {\bibinfo  {journal} {Phys.
  Rev. D}\ }\textbf {\bibinfo {volume} {84}},\ \bibinfo {pages} {125016}
  (\bibinfo {year} {2011})}\BibitemShut {NoStop}%
\bibitem [{\citenamefont {Gerdt}\ \emph
  {et~al.}(2011{\natexlab{b}})\citenamefont {Gerdt}, \citenamefont
  {Khvedelidze},\ and\ \citenamefont {Palii}}]{GKP2}%
  \BibitemOpen
  \bibfield  {author} {\bibinfo {author} {\bibfnamefont {V.}~\bibnamefont
  {Gerdt}}, \bibinfo {author} {\bibfnamefont {A.}~\bibnamefont {Khvedelidze}},
  \ and\ \bibinfo {author} {\bibfnamefont {Y.}~\bibnamefont {Palii}},\
  }\href@noop {} {\bibfield  {journal} {\bibinfo  {journal} {Phys. Atomic
  Nuclei}\ }\textbf {\bibinfo {volume} {74}},\ \bibinfo {pages} {893} (\bibinfo
  {year} {2011}{\natexlab{b}})}\BibitemShut {NoStop}%
\bibitem [{\citenamefont {Dunkl}\ and\ \citenamefont
  {Slater}(2015)}]{LatestCollaboration2}%
  \BibitemOpen
  \bibfield  {author} {\bibinfo {author} {\bibfnamefont {C.~F.}\ \bibnamefont
  {Dunkl}}\ and\ \bibinfo {author} {\bibfnamefont {P.~B.}\ \bibnamefont
  {Slater}},\ }\href@noop {} {\bibfield  {journal} {\bibinfo  {journal} {Random
  Matrices Theory Appl.}\ }\textbf {\bibinfo {volume} {4}},\ \bibinfo {pages}
  {1550018} (\bibinfo {year} {2015})}\BibitemShut {NoStop}%
\bibitem [{\citenamefont {Aubrun}\ \emph {et~al.}(2014)\citenamefont {Aubrun},
  \citenamefont {Szarek},\ and\ \citenamefont {Ye}}]{aubrun2}%
  \BibitemOpen
  \bibfield  {author} {\bibinfo {author} {\bibfnamefont {G.}~\bibnamefont
  {Aubrun}}, \bibinfo {author} {\bibfnamefont {S.~J.}\ \bibnamefont {Szarek}},
  \ and\ \bibinfo {author} {\bibfnamefont {D.}~\bibnamefont {Ye}},\ }\href@noop
  {} {\bibfield  {journal} {\bibinfo  {journal} {Commun. Pure Appl. Math.}\
  }\textbf {\bibinfo {volume} {LXVII}},\ \bibinfo {pages} {0129} (\bibinfo
  {year} {2014})}\BibitemShut {NoStop}%
\bibitem [{\citenamefont {Sommers}\ and\ \citenamefont
  {{\.Z}yczkowski}(2003)}]{szBures}%
  \BibitemOpen
  \bibfield  {author} {\bibinfo {author} {\bibfnamefont {H.-J.}\ \bibnamefont
  {Sommers}}\ and\ \bibinfo {author} {\bibfnamefont {K.}~\bibnamefont
  {{\.Z}yczkowski}},\ }\href@noop {} {\bibfield  {journal} {\bibinfo  {journal}
  {J. Phys. A}\ }\textbf {\bibinfo {volume} {36}},\ \bibinfo {pages} {10083}
  (\bibinfo {year} {2003})}\BibitemShut {NoStop}%
\bibitem [{\citenamefont {Altafini}(2004)}]{Altafini}%
  \BibitemOpen
  \bibfield  {author} {\bibinfo {author} {\bibfnamefont {C.}~\bibnamefont
  {Altafini}},\ }\href@noop {} {\bibfield  {journal} {\bibinfo  {journal}
  {Phys. Rev. A}\ }\textbf {\bibinfo {volume} {69}},\ \bibinfo {pages} {012311}
  (\bibinfo {year} {2004})}\BibitemShut {NoStop}%
\end{thebibliography}%

\end{document}